\documentclass[12pt]{article} 
\usepackage{amssymb}
\usepackage{graphicx}
\begin{document} 
\begin{center}
{\large \bf Will protons become gray at 13 TeV and 100 TeV?}

\vspace{0.5cm}                   

 I.M. Dremin

\vspace{0.3cm}                       

         Lebedev Physical Institute, Moscow 119991, Russia

\vspace{0.3cm}

National Research Nuclear University "MEPhI", Moscow 115409, Russia     

\end{center}

\begin{abstract}
It is shown that the regime of pp-interactions at 7 TeV is a critical one.
The LHC data about elastic pp-scattering at 7 and 8 TeV are used to get
some information about both elastic and inelastic profiles of pp-collisions.
They are discussed in the context of two phenomenological models which intend
to describe the high energy pp-data with high accuracy. Some predictions
following from these models for LHC energy 13 TeV and for energy 95 TeV
of the newly proposed collider are discussed. It is claimed that the center of 
the inelastic interaction region will become less dark with increase of energy
albeit very slowly. 
\end{abstract}

\section{Introduction}

The data of TOTEM collaboration \cite{tot7, tot8} about elastic scattering
of protons at energies 7 and 8 TeV revived interest to its characteristics
(for recent reviews see \cite{dr13, god}).  They are important not only by
themselves but also for getting some information about inelastic processes.
One of the most exciting findings from the data is the completely dark profile 
of central inelastic collisions \cite{sfr, dnec}. The knowledge about it has 
been obtained directly from the unitarity condition. It states that the total 
probability of all outcomes of proton collisions must be equal 1 at a given 
energy and relates the elastic and inelastic amplitudes. Inserting there the 
experimental data about elastic scattering one gets some conclusions about
inelastic events. The saturation limit of the inelastic profile equal 1
implies the complete darkness of the interaction region.

\section{Profiles}

By the profile of collisions we mean the impact parameter b distribution
of the strength of the interaction. The impact parameter is defined as the 
shortest transverse distance between the trajectories of the protons' centers.
In this representation the unitarity condition looks as
\begin{equation}             
G(s,b)=2\Gamma (s,b)- \Gamma ^2(s,b).
\label{unit1}
\end{equation}
$\Gamma (s,b)$ is defined as the Fourier - Bessel transform of the elastic
scattering amplitude $f(s,t)$ which depends on energy s and transferred 
momentum t. It retranslates the transferred momentum data to the transverse 
impact parameter space features and is written as
\begin{equation}
\Gamma (s,b) \approx \frac {1}{2\sqrt {\pi }}\int _0^{\infty}d\vert t\vert 
{\rm Im}f(s,t) J_0(b\sqrt {\vert t\vert }).
\label{gamm}
\end{equation}
The real part of the amplitude is small and will be neglected so that
${\rm Im}f(s,t)\approx \sqrt {d\sigma /dt}$.

The left-hand side of Eq. (\ref{unit1}) describes the impact parameter profile 
of inelastic 
collisions of protons. It satisfies the inequalities $0\leq G(s,b)\leq 1$ and 
determines how absorptive is the inelastic interaction region depending on the 
impact parameter (with $G=1$ for the full absorption and $G=0$ for the complete 
transparency). The profile of elastic processes is determined by the subtrahend 
in Eq. (\ref{unit1}). If $G(s,b)$ is integrated over the impact parameter, it 
leads to the cross section of inelastic processes. The terms on the right-hand 
side would produce the total cross section and the elastic cross section. 

The elastic differential cross section is especially large at small transferred
momenta within the so-called diffraction cone where it decreases exponentially
with $\vert t\vert $. Thus, the diffraction cone contributes mostly to the 
Fourier - Bessel transform of the amplitude. Using the above formulae, one 
can write the dimensionless $\Gamma $ as
\begin{equation}
\Gamma (s,b)=\frac {\sigma _{tot}}{8\pi }\int _0^{\infty}d\vert t\vert 
\exp (-B\vert t\vert /2 )J_0(b\sqrt {\vert t\vert }).
\label{gam2}
\end{equation}
Here, the diffraction cone approximation is inserted: 
\begin{equation}
\frac {d\sigma }{dt}=\vert f(s,t)\vert ^2 \approx
\frac {\sigma ^2_{tot}}{16\pi }\exp (-B\vert t\vert ).
\label{expB}
\end{equation}

Herefrom, one calculates
\begin{equation}
\Gamma (s,b)=\zeta \exp (-\frac {b^2}{2B}),
\label{rega}
\end{equation}
where we introduce the dimensionless ratio of the cone slope (or the elastic
cross section) to the total cross section
\begin{equation}
\zeta =\frac {\sigma _{tot}}{4\pi B}\approx \frac {4\sigma _{el}}{\sigma _{tot}}.
\label{ze}
\end{equation}
Possible small deviations from the exponential behavior inside the cone do not 
practically change the value of the integral contribution.
The differential cross section is quite small outside the diffraction peak
and does not influence the impact parameter profile $G$. 

Thus, we get
\begin{equation}
G(s,b)= \zeta \exp (-\frac {b^2}{2B})(2-\zeta \exp (-\frac {b^2}{2B})).
\label{ge}
\end{equation}
The inelastic profile $G(s,b)$ is directly obtained from the unitarity condition
as a difference of two Gaussian exponentials. The exponents differ by a factor 
of two. The shapes of the profiles are completely determined by the slope B(s)
and the total cross section $\sigma _{tot}(s)$. They have been computed at
different energies using the corresponding experimental data. From ISR 
(62.5 GeV) to LHC (7 TeV) energies the inelastic interaction region of protons
becomes darker and extends to larger impact parameters. That supports earlier
expectations that the protons become more black, edgier and larger (BEL). At the 
LHC energy 7 TeV the inelastic profile is extremely dark up to about 0.5 fm
and its width is about 1.5 fm while at the ISR energy 62.5 GeV the darkness was 
slightly above 0.9 at the center and decreased twice already at about 0.7 fm.

The elastic profile given by the subtrahend in (\ref{ge}) is comparatively
more narrow. The inelastic processes are always more peripheral than the 
elastic ones. The ratio of the average values of the squared impact parameters 
is given by
\begin{equation}
\frac {<b^2_{inel}>}{<b^2_{el}>}=\frac {8-\zeta }{4-\zeta }
\label{b2}       
\end{equation}
as is easily seen from their profiles normalized by the corresponding cross
sections. 

For central collisions with $b=0$ one gets
\begin{equation}
G(s,b=0)=\zeta (2-\zeta ).
\label{gZ}       
\end{equation}
This formula is very important. Herefrom, it follows that the darkness at the 
very center $b=0$ is fully determined by the single parameter $\zeta $, 
i.e. by the ratio of experimentally measured characteristics - the width of the 
diffraction cone $B$ (or $\sigma _{el}$) to the total cross section.
Its energy evolution defines the evolution of the absorption value.

As a function of energy, the value of $\zeta $ changes in the range of ISR to 
LHC from 0.66 to 1.0 (with intermediate values of 0.8 at S$\bar pp$S at 546 GeV 
and 0.9 at Tevatron at 1.8 TeV if the proton-antiproton data are included).
Therefore, compared to ISR data, the darkness of central collisions strongly 
increases at the LHC. The inelastic interaction region becomes completely 
black at impact parameters $b$ up to 0.5 fm \cite{dr15}. 
The interaction region is completely absorptive ($G(s,0)=1$) in the center 
only at $\zeta =1$ and the absorption diminishes for other values of $\zeta $. 
Eq. (\ref{gZ}) shows that the darkness at the center changes extremely slowly
for values of $\zeta $ close to 1. The decline by $\pm \epsilon $ from 1 in 
$\zeta $ leads to corrections of the order of $\epsilon ^2$ in 
$G(s,b=0)=1-\epsilon ^2$. Therefore, the regime of pp-interactions at 7 TeV 
can be considered as a critical one.

That has some important consequences for inelastic processes
indicating that high density gluon configurations in the 
interaction region play a crucial role for high multiplicity events at LHC 
energies \cite{ads}. For example, it was noticed at the
CMS studies that experimental results at 7 TeV on jet production in rare events
with the very high charged particle multiplicities above 70 differ 
from the predictions of the widely used Monte Carlo models PYTHIA and HERWIG. 
The production of jets is a typical feature of such events.
The jets are mostly produced at central collisions where the density of partons 
(or strings) is high so that multiparton interactions become important and the 
interactions themselves are strong enough. It has been shown that at the very 
high multiplicities the rate of jet production is not described by the geometry 
of pp-interactions. These processes become dominated by rare central collisions,
where some special high-density gluon fluctuations appear inside colliding 
protons. Therefore it looks likely that one should include more dense 
gluon states in Monte Carlo models to get an agreement with experiment.

\section{Expectations}

What can we expect at higher energies?

The only guesses can be obtained from the extrapolation of results at lower 
energies to new regimes even though our experience shows how indefinite and 
even erroneous they can be as it often happened.

First, one may assume that $\zeta $ will increase without crossing 1 but 
approaching it asymptotically. That would imply that its precise value at 7 TeV 
is still slightly lower than 1 within the present experimental errors. Then the 
inelastic profile will be quite stable with slow approach to the complete
blackness at central collisions and steady increase of its range. That is
a kind of "the black stick" if one implies rather long longitudinal distances
as it is commonly believed for the picture with wee partons.

Another, more intriguing possibility, discussed below in the framework of some 
phenomenological models, is the further increase of $\zeta $ above 1. Then the 
darkness at the very central collisions $G(s,b=0)$ diminishes in accordance with
Eq. (\ref{gZ}). The center becomes more transparent. The dip should appear 
inside the plateau with a minimum value at $b=0$. The black bump at the center
observed at 7 TeV transforms to the toroid-like structure with somewhat lower 
darkness at the center and maximum blackness equal 1 at some peripheral 
impact parameter $b_m$. This dependence is very slow near $\zeta =1$ so that the 
darkness at the center would only become smaller by 6$\%$ if $\zeta $ increases 
to 1.2. Therefore one can hardly expect the immediate drastic changes with 
increase of LHC energies. 
Nevertheless, the forthcoming TOTEM+CMS results on elastic scattering at 13 TeV 
can be very conclusive about the general trend if the precise values of the 
diffraction cone slope B and the total cross section $\sigma _{tot}$ become 
available and the corresponding value of $\zeta $ happens to exceed 1.

This tendency is supported by recent phenomenological models \cite{kfk, fms}. 
In one of them (KFK), the detailed form of the elastic scattering amplitude
is prescribed. The diffraction cone region at small $\vert t\vert $ and further 
Orear-type regime at larger $\vert t\vert $ with exponential decrease in 
$\sqrt {\vert t\vert}$ are parameterized and successfully compared with present 
experimental data. By passing, we note that the real part of the amplitude 
computed in the model is extremely small within the diffraction cone and even 
passes zero inside it. That confirms our assumption about it. In another model 
(FMS), the energy behaviour of the ratio of the elastic and total 
cross sections is studied. According to Eq. (\ref{ze}), the energy behaviour 
of $\zeta $ has been directly modeled. Both models intend to describe quite well the 
presently available experimental data and give some predictions for higher 
energies. 

In the Table, the values of the slope B(s) and the total cross section 
$\sigma _{tot}(s)$ available in KFK are reproduced. We have calculated the 
corresponding values of $\zeta $ and $G(s,b=0)$ for KFK extending the range of 
energies up to its asymptotics. The values of $\zeta $ and $G(s,b=0)$ have also 
been computed for FMS. It is clearly seen that both models favor slight decrease 
of the inelastic profile at the center with the shift of maximum blackness 
equal 1 to peripheral impact parameters $b_m(s)$ which are shown in a separate 
column.
 
\begin{table}
\medskip
Table.  $\;\;$ The energy behavior of $\zeta (s)$ and $G(s,b=0)$.
\medskip

    \begin{tabular}{|l|l|l|l|l|l|l|l|l|l|l|l}
        \hline
$        $&$\sqrt s$,TeV&$\zeta $(s)&G(s,0)&$b_m(s)$,fm&B(s),GeV$^{-2}$&$\sigma _{tot}(s)$, mb\\ \hline
$TOTEM  \cite{tot7}$   & 7   &1.0128 & 0.9998 & 0.14 & 19.9 & 98.6  \\
        \cite{tot8}  & 8  & 1.0189 & 0.9996 & 0.17 & 20.4 & 101.7  \\  \hline
$ KFK  \cite{kfk}$     & 7  & 1.0133 & 0.9998 & 0.14 & 19.9 & 98.65  \\  
$          $ & 8  & 1.0215 & 0.9995 & 0.18 & 20.21 & 101.00 \\
$           $  & 13  & 1.0524 & 0.9973 & 0.29 & 21.35 & 109.93  \\  \hline
$ computed  $  & 95  & 1.1497 & 0.9776 & 0.53 & 27.10 & 152.43  \\ 
$           $  & 1000 & 1.2393 & 0.9428 & 0.74 & 35.50 & 215.24  \\    
$           $  & $\infty $ & 1.4399 & 0.8060 & $\infty$ & $\infty $ & $\infty $ \\  \hline
$ FMS  \cite{fms}  $  & 8  & 1.0300 & 0.9991 \\
$              $  & 13  & 1.0554 & 0.9969   \\
$           $  & 95  & 1.1231 & 0.9848 \\  \hline   
   	
\end{tabular}
\end{table}
Both models predict slightly faster increase of $\zeta $ with energy from 7 to 
8 TeV if compared to the TOTEM data albeit within experimental errors. The 
values of $G(s,0)$ do not change, in practice, even though slight tendency to 
decrease in the fourth (!) digit is shown in the Table. At the energy 13 TeV 
the value of $\zeta $ differs from that at 7 TeV only less than by 4$\% $. 
Correspondingly, the darkness is different at the third digit. That puts some 
hard problems for experimenting at 13 TeV. Nevertheless, new experimental data 
at 13 TeV would show the general qualitative trend if their accuracy is high 
enough.

What concerns higher energies, the above tendency persists. The central region
becomes more gray. It may be noticed in the data from the newly proposed 100 
TeV-collider as shown in the Table in rows at 95 TeV if one trusts the models'
predictions. Due to the assumptions of some saturation in asymptotics, used in 
the models, the predictions evolve extremely slowly with energy so that even 
the asymptotical value of $G(s,0)$ is close to 0.8 as compared to the 
absolute blackness 1 at peripheral impact parameters $b_m(s)$ also shown in the
Table. 

It is hard to foresee any special features of inelastic processes if the 
evolution will be that slow.  The (almost) black plateau will become somewhat
enlarged in size. Therefore the jet cross sections due to central collisions 
will increase as well. The inelastic profile will become even more peripheral
and the role of peripheral collision will increase.  

In principle, this tendency will be preserved until the value of $\zeta $ 
becomes equal to 2 which erroneously is often called as the black disk regime. 
Then at the corresponding energy $s_0$ the center of the inelastic 
interaction region becomes completely transparent $G(s_0,b=0)=0.$ It means that
elastic processes will only dominate at the very central collisions, and
the inelastic events will originate at more peripheral regions. Surely, this
possibility looks quite opposite to our intuition. Nevertheless, it can
happen according to the above dependence of the central darkness $G(s,b=0)$
on $\zeta $. No model predicts the fast rise of $\zeta $ to values close to 2. 
Their preferred asymptotic values of $\zeta $ are less than 1.5 as shown in the 
Table. 

The values of $\zeta $ larger than 2 would correspond to another branch of the 
solution of the unitarity equation which describes the backward scattering 
(the reflective mode). Its realization in the near future looks very improbable 
even in the models and we do not consider it here.

To conclude, the latest findings of TOTEM and CMS  contributed much to 
understanding of the shape of the interaction region of colliding protons.
Its structure changes with increase of their energies and one can await that 
some new features will be noticed at higher energies.

I am grateful for support by the RFBR grant 14-02-00099 and by the RAS-CERN 
program.

\end{document}